# Pervasive symmetry-lowering nanoscale structural fluctuations in the cuprate La$_{2-x}$Sr$_x$CuO$_4$

R. J. Spieker[1], M. Spaić[2], I. Khayr[1], X. He[1], D. Zhai[1], Z. W. Anderson[1,&], N. Bielinski[1,#], F. Ye[3], Y. Liu[3,4], S. Chi[3], S. Sarker[5], M. J. Krogstad[6] R. Osborn[7], D. Pelc[1,2*], and M. Greven[1*]

[1] School of Physics and Astronomy, University of Minnesota, Minneapolis, MN, USA

[2] Department of Physics, Faculty of Science, University of Zagreb, Bijenička 32, 10000 Zagreb, Croatia

[3] Neutron Scattering Division, Oak Ridge National Laboratory, Oak Ridge, TN, USA

[4] Second Target Station, Oak Ridge National Laboratory, Oak Ridge, TN, USA

[5] Cornell High Energy Synchrotron Source, Cornell University, Ithaca, NY, USA

[6] Advanced Photon Source, Argonne National Laboratory, Lemont, IL, 60439

[7] Materials Science Division, Argonne National Laboratory, Argonne, IL, USA

[&] Present address: Materials Science Division, Argonne National Laboratory, Argonne, IL, USA

[#] Present address: Department of Physics, University of Illinois at Urbana-Champaign, Champaign, IL, USA

*Correspondence to: dpelc@phy.hr, greven@umn.edu,

The cuprate superconductors are among the most widely studied quantum materials, yet there remain fundamental open questions regarding their electronic properties and the role of the structural degrees of freedom. Recent neutron and x-ray scattering measurements uncovered exponential scaling with temperature of the strength of orthorhombic fluctuations in the tetragonal phase of La$_{2-x}$Sr$_x$CuO$_4$ and Tl$_2$Ba$_2$CuO$_{6+\delta}$, unusual behavior that closely resembles prior results for the emergence of superconducting fluctuations, and that points to a common origin rooted in inherent correlated structural inhomogeneity. Here we extend the measurements of La$_{2-x}$Sr$_x$CuO$_4$ to higher temperatures in the parent compound ($x = 0$) and to optimal doping ($x = 0.155$), and we furthermore investigate the effects of *in-situ* in-plane uniaxial stress. Our neutron scattering result for undoped La$_2$CuO$_4$ complement prior x-ray data and reveal that the structural fluctuations persist to the maximum experimental temperature of nearly 1000 K, *i.e.*, to a significant fraction of the melting point. At this temperature, the spatial correlation length extracted from the momentum-space data is still about three lattice constants. The neutron scattering experiment enables quasistatic discrimination and reveals that the response is increasingly dynamic at higher temperatures. We also find that uniaxial stress up to 500 MPa along the tetragonal [1 1 0] direction, which corresponds to a strain of about 0.2%, does not significantly alter this robust behavior. Overall, these results support the notion that subtle, underlying inhomogeneity underpins the cuprate phase diagram. Finally, we uncover (for $x = 0.20$) low-energy structural fluctuations at a nominally forbidden reflection. While the origin of these fluctuations is not clear, they might be related to the presence of extended defects such as dislocations or stacking faults.

## I. Introduction

The cuprates constitute a large family of lamellar perovskite-derived oxides with a rich phase diagram that features 'strange-metal' charge transport, pseudogap behavior, as well as superconductivity with record high ambient-pressure transition temperatures ($T_c$) [1]. Experimental evidence indicates a prevalence of electronic and structural inhomogeneity [2-14], yet pivotal aspects of the connection between structural and electronic degrees of freedom remain unresolved. Importantly, the emergence of superconducting fluctuations is highly unusual, as it is characterized by an extended exponential temperature dependence and scaling with relative temperature, $T - T_c$, in numerous observables [11-13]. This robust behavior is qualitatively different from conventional fluctuations, *i.e.*, a power-law dependence on reduced temperature, $(T - T_c)/T_c$. An exponential temperature dependence has also been reported for the structurally related lamellar ruthenate $Sr_2RuO_4$ and the classic perovskite $SrTiO_3$ and proposed to be associated with intrinsic, self-organized structural inhomogeneity inherent to the oxides' perovskite-based structure [14].

Some cuprates, such as $La_{2-x}Sr_xCuO_4$ (LSCO) and $Tl_2Ba_2CuO_{6+\delta}$, undergo a structural transition from a high-temperature tetragonal (HTT) to a low-temperature orthorhombic (LTO) phase. Interestingly, exponential scaling with relative temperature was recently observed for the strength of LTO diffuse scattering in the HTT phase [15], closely similar to the superconducting fluctuation behavior. Such scaling can generically arise from rare-region effects, whereby exponentially rare, ordered puddles form above the bulk phase transition temperature due to underlying inhomogeneity. We emphasize that, in this picture, the observed unusual LTO diffuse scattering is a non-universal signature of underlying inhomogeneity, as most cuprates do not exhibit a tendency toward LTO distortions. The closely similar behavior exhibited by the electronic and structural degrees of freedom therefore provides evidence for inherent correlated structural inhomogeneity in the cuprates and other complex oxides, and it suggests that the respective order parameters couple to local strain. In the case of the cuprates, these observations furthermore support a recent phenomenological model in which the evolution with doping from the undoped Mott insulator to the Fermi liquid involves a second, thermally-activated carrier component with a broad distribution of local activation gaps [14,15].

Here we perform neutron and x-ray diffuse scattering measurements of LSCO to expand on the recent experiments [15] and to gain more information about the structural degrees of freedom in the nominally tetragonal phase of the undoped Mott insulator ($x = 0$) and the optimally doped system ($x = 0.155$). The structural phase transition temperature of LSCO, $T_{LTO}$, decreases approximately linearly with doping [16] (Fig. 1). This transition involves the rigid staggered rotation of the $CuO_6$ octahedra about the in-plane diagonals $[1\ 1\ 0]$ and $[1\ \bar{1}\ 0]$ (we use tetragonal notation throughout) and results in macroscopic twin domains. It has been known for some time that LTO fluctuations are present in the HTT phase [17-20], including at high Sr concentrations, where the average structure remains tetragonal to the lowest measured temperatures [17,20]. However, the unusual exponential temperature dependence and doping-independent scaling with relative temperature were reported only recently [15]. Our new neutron scattering data reveal that the predominantly dynamic structural fluctuations persist to at least 970 K in the stoichiometric parent compound $La_2CuO_4$, which indicates that they are unrelated to substitutional disorder and likely emerge near the melting point of the material. The relevant energy scale for the underlying structural inhomogeneity is thus quite high and comparable to key electronic scales, such as the antiferromagnetic superexchange of the parent compounds (~130 meV) [21] and the superconducting gap (~50 meV at optimal doping) [22,23], with significant implications for electronic models of the cuprate phase diagram that are rooted in considerations of inhomogeneity [24,25]. At optimal doping, our combined neutron and x-ray data are consistent with the prior results at lower and higher doping levels, yet the correlation length is about 50% smaller at the relative temperatures $T - T_{LTO}$. This effect is seen in multiple samples and might be the result of subtle changes in the lattice properties around optimal doping. We do not observe significant qualitative changes with *in situ* uniaxial stress along [1 1 0], which further demonstrates the robustness of the unusual fluctuation behavior. Finally, we consider the structural response at reflections that are nominally forbidden in the LTO phase. Although the average low-temperature structure of LSCO is generally thought to be orthorhombic at doping levels below $x = 0.22$ [16], Bragg peaks consistent with monoclinic distortions have been observed for $x = 0$ (hk0 for half integer h,k) and $x = 0.05$ (hk0 for half integer h,k and hhl for half-integer h with odd l) [26,27]. In the present work, we find that an overdoped ($x = 0.20$) sample exhibits weak quasistatic peaks at $(0\ \bar{3}\ 4)$ and equivalent positions, in addition to a previously observed diffuse feature. While the peaks may be a bulk effect indicative of an average structure that is lower than orthorhombic, another possibility is that the signal originates from extended defects such as domain walls or stacking faults.

This paper is organized as follows: Section II describes the experimental methods employed in this work - crystal growth and characterization as well as details pertaining to the x-ray and neutron scattering measurements. In Section III, we present the results of our scattering experiments. Finally, in Section IV, we explore the implications of these findings and summarize our results.

## II. Experimental Methods

*Crystal Growth and Characterization.* Single crystals of undoped ($x = 0$), optimally doped (nominal Sr content $x = 0.155$), and overdoped (nominally $x = 0.20$) La$_{2-x}$Sr$_x$CuO$_4$ were grown at the University of Minnesota with the travelling-solvent floating-zone (TSFZ) method. The undoped and $x = 0.20$ samples were measured *via* diffuse neutron scattering. For the latter sample, the short-range LTO correlations were reported in [15], and an additional diffuse feature was observed at nominally forbidden lower-symmetry reflections such as $(3\ 0\ \bar{4})$ (see the *Supplementary Material* of [15]). Three optimally-doped samples were measured: two *via* diffuse x-ray scattering (one under ambient conditions and one under uniaxial strain), and the third *via* triple-axis neutron scattering. The latter sample was used in previous work to demonstrate the existence of local LTO modes in the tetragonal phase using triple-axis spectroscopy [15]; here we report the temperature dependence of the LTO fluctuations. All other compositions and data shown in Figs. 4 and 5b are from [15].

The Sr-doped samples were annealed at 800°C in O$_2$ to relieve internal stress due to the growth process. The nominally undoped sample was reduced in flowing argon at 800°C to remove excess oxygen and, hence, hole carriers [28,29]. Superconducting samples were characterized by nonlinear (third-harmonic) AC magnetic susceptibility measurements [14], which confirmed the presence of sharp superconducting transitions. The $T_c$ values were consistent with hole concentrations that match the nominal Sr concentrations [30]. For the nominally undoped sample, a standard DC magnetic susceptibility measurement showed the expected antiferromagnetic transition, with a high Néel temperature of $T_N \approx 325$ K, indicative of the absence of holes [28,29]. The sample masses were different for each scattering experiment, and the crystals were cut/polished as needed. The samples studied *via* x-ray scattering had masses of 1 to 10 mg. The $x = 0.155$ sample used for the *in-situ* uniaxial strain measurement was polished into a cuboid shape with surface roughness of 0.1 μm with sides corresponding to the [1 1 0], [1 $\bar{1}$ 0], and [0 0 1] high-symmetry directions. Stress of up to 500 MPa was applied along [1 1 0]. This strain direction was selected because it couples strongly to the CuO$_6$ rotation of the LTO phase, where modest stresses of only ~20 MPa are sufficient to create a single-domain sample [31]. Using an in-pane elastic modulus of ~250 GPa [32], the maximum applied stress of 500 MPa corresponds to a strain of about 0.2%. Samples used in the diffuse neutron scattering measurements had masses of about 0.5 g. The triple-axis neutron scattering sample consisted of two co-mounted crystals from a single TSFZ growth with a combined mass of about 7.5 g and a Bragg-peak mosaic spread of less than 0.6° (FWHM).

*Diffuse X-ray and Neutron Scattering*. Diffuse x-ray scattering measurements were carried out at beamline 6-ID-D of the Advanced Photon Source (APS) at Argonne National Laboratory and beamline ID4B (now QM2) at the Cornell High Energy Synchrotron Source (CHESS). This approach uses high-energy x-rays to capture the energy-integrated response; the energies were 87 keV and 37 keV at APS and CHESS, respectively. A high photon flux and millimeter-sized beam spots enabled the investigation of sub-millimeter sized samples with short measurement times, allowing for data collection at many temperatures. Data were collected during φ rotations through 360° of the single crystals using a fast CdTe detector and a Pilatus 6M detector at the APS and CHESS, respectively. An Oxford Cryostream was used to cool samples at both the APS and CHESS (see also the Supplementary Information in [33]). In the *in situ* uniaxial strain experiment, we used a dedicated pneumatic uniaxial strain cell that is compatible with the flowing helium gas cooler at beamline 6-ID-D at the APS. The sample was mounted between a steel piston and topaz anvil, and high-pressure helium gas was used to drive the piston and apply stress to the sample, similar to previous designs [14,34,35]. This novel device allows for significantly larger compressive stress values (up to 200 N) than existing piezoelectric cells of similar size [36-38], with high stress homogeneity due to the absence of any glue. Measurements with the cell were performed in a limited φ-range and with zero ω-tilt due to geometrical constraints and hence show a lower signal-to-noise ratio than full rotation/tilt scans (typically, data were collected for three sets of rotations, with the axis tilted by $\chi = -15°, 0°,$ and $15°$). Diffuse neutron scattering measurements were performed with the CORELLI instrument at the Spallation Neutron Source at Oak Ridge National Laboratory. CORELLI uses a high-flux white neutron beam in conjunction with state-of-the-art statistical choppers that allow for the discrimination of the quasistatic signal ($\lesssim 2$ meV) from the "total" energy-integrated scattering ($\lesssim 10$ meV). A top-loading furnace capable of reaching 1,000 K was used to heat the $x = 0$ sample, and the expected transition temperature $T_{\text{LTO}} = 530$ K was assumed when displaying data vs. relative temperature, $T - T_{\text{LTO}}$ [28].

The superstructure intensities in Figs. 2, 4, and 5 were obtained by integrating over three-dimensional boxes in reciprocal space with sides of lengths 0.3-0.6 r.l.u. centered on the respective LTO peaks. Background correction was performed by subtracting the integrated intensity from a nearby box with the same reciprocal space volume. The LTO positions in Figs. 2 and 4 are (2.5 1.5 9) for $x = 0$ (neutron) and $(\overline{2.5}\ \overline{2.5}\ 2)$ for $x = 0.155$ (x-ray); as noted, we use tetragonal notation throughout. For the strain-cell experiment, the superstructure intensity shown in Fig. 5(a) was taken to be the average of the response at five LTO positions: $(4.5\ \overline{1.5}\ 9)$, $(5.5\ \overline{2.5}\ 9)$, $(6.5\ \overline{1.5}\ 9)$, $(6.5\ \overline{3.5}\ 9)$, and $(7.5\ \overline{2.5}\ 9)$. The corresponding Gaussian widths (Fig. 5(b)) were determined from the average width of three LTO positions: $(4.5\ \overline{1.5}\ 9)$, $(6.5\ \overline{1.5}\ 9)$, and $(7.5\ \overline{2.5}\ 9)$. This method was used because of the relatively large noise in the signal, which may have resulted from quick single-axis measurements, as noted above. Random thermal atomic displacements in general only affect the observed diffuse scattering intensities through the Debye-Waller factor in the cross section, which has the approximate exponential form $\exp(-Q^2 u^2)$, where $Q$ and $u$ are the wavevector magnitude and average thermal displacement, respectively. The Debye-Waller factor contribution to the temperature dependences of the integrated diffuse intensities was investigated elsewhere and determined to be negligible [15]. The widths of the diffuse LTO features were extracted from the neutron and x-ray diffuse scattering data by fitting to a one-dimensional Gaussian profile plus a third-order polynomial background after integrating out the remaining two directions. Gaussian functions were found to be a better fit than Lorentzian profiles at all temperatures, but similar widths are obtained from fits with the latter. The correlation lengths were determined from the half-width-at-half-maximum (HWHM) using the method in [15].

*Triple-Axis Neutron Spectroscopy.* The triple-axis neutron scattering measurements were performed with the thermal spectrometer HB-3 at the High-Flux Isotope Reactor, Oak Ridge National Laboratory. Standard collimation of 48' – 40' – sample – 40' – 120' was used, with a fixed final energy of 14.7 meV. The full-width half-maximum (FWHM) energy resolution was estimated to be 1.3 meV at the elastic line. $[H\ H\ 0]$ scans were performed through the (1.5 1.5 2) LTO position.

**III. Results**

The prior diffuse x-ray scattering measurements of the structural correlations in undoped $La_2CuO_4$ extended to $T$ = 700 K [15]. In Fig. 2, we present our new neutron scattering results for $x$ = 0, which complement these x-ray data and extend to considerably higher temperatures. The diffuse LTO features are asymmetric along $[1\ 1\ 0]$ and $[1\ \bar{1}\ 0]$ (Fig. 2(a)), consistent with previous observations for undoped and underdoped LSCO (see the *Supplementary Material* of [15]). Figure 2(b) shows momentum scans along $[1\ 1\ 0]$ and $[1\ \bar{1}\ 0]$ at 630 K. LTO fluctuations are clearly visible in the HTT phase in both the quasistatic and total scattering channels, and in both cases the temperature dependence is exponential, $I \sim e^{-T/T_0}$ (Fig. 2(c)). The slope of the quasistatic scattering ($T_0$ = 179 ± 26 K) is somewhat steeper than that of the total scattering ($T_0$ = 286 ± 25 K), which indicates that the response at higher temperatures is increasingly dominated by dynamic fluctuations. Nevertheless, the nonzero quasistatic response indicates that LTO correlations are "frozen-in" even well above the bulk transition. The characteristic temperature scales are comparable to the prior x-ray result $T_0$ = 155 ± 8 K for $La_2CuO_4$ (see also Fig. 4(a)). Intriguingly, the exponential behavior is observed up to the maximum experimental temperature of 970 K, which is about 60% of the melting point of approximately 1600 K [39]. The observed LTO peaks are commensurate in both the quasistatic and total scattering channels, as expected for the undoped parent compound [17]. The peak widths are much larger than those of resolution-limited neighboring Bragg peaks, indicative of short-range correlations. The correlation lengths determined from the average of the Gaussian widths of the total scattering response along $[1\ 1\ 0]$ and $[1\ \bar{1}\ 0]$ exhibit a power-law temperature dependence (Fig. 2(d)), consistent with prior observations [15]. In the case of the quasistatic response, correlation lengths could only be extracted reliably below 750 K due to the relatively low signal-to-noise ratios at higher temperatures. The length extracted from the quasistatic response (effective energy integration up to 1-2 meV) increases somewhat faster on cooling that that extracted from the total scattering (integration up to ~ 10 meV).

The LTO peaks are asymmetric, as previously reported [15], with different characteristic length scales along $[1\ 1\ 0]$ and $[1\ \bar{1}\ 0]$. Figure 3(a,b) compares the (5/2 3/2 9) and (7/2 7/2 2) peaks. The results for the (5/2 3/2 9) peak are shown in Fig. 2. The (7/2 7/2 2) peak was selected for comparison since this momentum transfer is primarily along $[1\ 1\ 0]$, which corresponds to one of the two LTO tilt directions (and, consequently, the orthorhombic $[1\ 0\ 0]$ direction). The correlation lengths extracted from the data in the two Brillouin zones are consistent with each other (Fig. 3(a)). However, the asymmetry, characterized by the ratio $\sigma_{[1\bar{1}0]}/\sigma_{[110]}$ of the Gaussian widths along the two orthogonal planar directions shows subtle differences. While $\sigma_{[1\bar{1}0]}/\sigma_{[110]} \approx 2$ for the (5/2 3/2 9) peak at all temperatures, the (7/2 7/2 2) data appear to approach an isotropic response for $T$ - $T_{LTO}$ > 250 K. The latter is likely a systematic error, as the signal becomes relatively weak at higher temperatures.

Figure 4(a) compares the temperature dependences of the diffuse LTO intensity for undoped and optimally doped $La_{2-x}Sr_xCuO_4$ with the prior results from ref. [15]. The small optimally-doped sample measured *via* x-ray scattering undergoes a sharp transition at $T_{LTO}$ ~ 150 K indicative of macroscopic sample homogeneity, whereas the optimally-doped sample measured *via* triple-axis neutron scattering features some macroscopic La/Sr inhomogeneity due to its large volume, which resulted in a somewhat broadened structural transition. Despite this difference, the characteristic scales $T_0$ of the observed exponential temperature dependences are consistent with the behavior observed over the entire doping range. As shown in Fig. 4(b), the correlation lengths for both undoped and optimally doped samples follow the previously established power-law dependence on relative temperature, $T - T_{LTO}$, for both the neutron and x-ray scattering data.

In a dedicated synchrotron x-ray scattering experiment, we investigated the effects of *in situ* uniaxial stress on the short-range LTO fluctuations in an optimally-doped sample. The crystal was compressively stressed from 25 MPa to 0.5 GPa along [1 1 0] at 300 K, in the HTT phase, and the diffuse scattering was measured on cooling. This stress direction was chosen because it enables the manipulation of the orthorhombic twin structure. In undeformed samples, the broken-symmetry phase consists of a twinned structure of macroscopic LTO domains with octahedral tilts about either [1 1 0] or [1 $\bar{1}$ 0]. Prior work demonstrated that a single-domain state is obtained upon cooling through the transition with a modest applied stress of ≥ 20 MPa along either of these directions [31]. Thus, the LSCO sample in our experiment was likely biased toward a single LTO domain configuration. Interestingly, as shown in Fig. 5, the measured intensity shows contrasting behavior below and above the structural transition temperature. While the intensity of the superstructure peak below $T_{LTO}$ increases with increasing stress (by about a factor of two at 0.5 GPa), the strength of LTO fluctuations in the HTT phase decreases (by about 30-40% at 0.5 GPa). Long-range order is thus stabilized by the applied stress, while fluctuations are suppressed. However, within the uncertainty of the experiment, the slope of the exponential temperature dependence remains unchanged ($T_0 = 222 \pm 31$ K), and the extracted correlation lengths show no strain dependence above 25 MPa. However, even already at at 25 MPa stress, the correlation lengths are systematically smaller than those observed at ambient pressure, indicative of a strain effect at stress levels below 25 MPa. Since 25 MPa is sufficient to fully de-twin samples upon cooling into the orthorhombic phase, such a strain effect at low stress might be expected. Although the collected data are relatively sparse around $T_{LTO}$, it appears that a stress of 0.5 GPa shifts $T_{LTO}$ upwards by about 20 K. At a qualitative level, such an increase is expected based on the observed enhancement of the superstructure peak intensity in the ordered phase. Moreover, it was previously observed that the diffuse x-ray intensity (after normalization by the intensity of a nearby Bragg) follows universal behavior with relative temperature, $T - T_{LTO}$ [15], for a wide range of Sr concentrations, from $x = 0$ to 0.27. This scaling appears to be lost with uniaxial strain: $T_{LTO}$ shifts upward with increasing strain, while the intensity decreases. More detailed measurements to higher strain values would shed further light on this interesting observation.

LSCO is close to structural instabilities other than the HTT-LTO transition. The original superconducting cuprate $La_{2-x}Ba_xCuO_4$ [40] exhibits nearly the same structural and electronic phase diagram as LSCO, but undergoes a structural phase transition near $x = 0.12$ from the LTO phase to a low-temperature tetragonal (LTT) phase at which superconductivity is severely suppressed [41,42]; the LTT distortion corresponds to a 45° rotation of the $CuO_6$ octahedra tilt axis within the CuO plane [42]. This additional instability is also seen in $(La,Nd)_{2-x}Sr_xCuO_4$ for sufficiently high Nd concentrations [43,44]. Interestingly, there have been reports of weak, nominally forbidden Bragg peaks in the LTO phase of undoped and heavily underdoped ($x = 0.05$) LSCO consistent with low-temperature tetragonal (LTT) and monoclinic-type distortions [26,27]. We carried out a diffuse neutron scattering measurement to search for such structural correlations at higher doping ($x = 0.20$). As shown in Fig. 6, we observe diffuse scattering around the forbidden (0 $\bar{3}$ 4) reflection. These fluctuations are largely dynamic in nature, with a small quasistatic component. Both components form rods along $L$ (Fig. 6(b,c)), indicative of predominantly two-dimensional correlations, and consistent with the presence of stacking faults or dislocations [45].

### IV. Discussion and summary

The data analysis in this work is simple in principle, but a few points merit discussion. In particular, the determination of correlation lengths from the diffuse LTO features requires assumptions about the nature of, and relationship between the static and dynamic components. The low-energy response is known to be incommensurate for strontium concentration above ~5%, while the features at lower doping levels are commensurate, as also seen in our $La_2CuO_4$ data [17]. Several explanations have been put forward to describe the incommensurate response. One of these considers modulated stripes of LTO domains separated by antiphase boundaries [17] and is inspired by the well-studied spin-stripe phenomena seen in $La_{2-x}Ba_xCuO_4$ and $La_{2-x-y}Nd_ySr_xCuO_4$ [43,46,47]. The other is an effective model that features an incommensurate static response with an antiphase boundary, along with a commensurate dynamic response, including a low-energy local mode [15]. Our focus here is on the unusual scaling behavior and does not require assumptions about microscopic properties, and we thus primarily analyze the width of the diffuse features. However, we do include a basic correlation length estimate for comparison. This length scale is extracted from the HWHM following the procedure in ref. [15]. As noted, we use Gaussian profiles, since they yield better fits to the data than Lorentzian; this might point to a distribution of intrinsic correlation lengths, which is not surprising given the intrinsic inhomogeneity of the short-range fluctuations. Our neutron scattering results for LCO ($x = 0$) indicate that the quasistatic correlation length is larger than that of the energy-integrated total scattering. Interestingly, the extracted correlation lengths for the uniaxially strained optimally-doped sample are systematically shorter than for unstrained samples (Fig. 5b). It is therefore possible that strain decreases the strength of static correlations relative to dynamic fluctuations, which could result in such a systematic shift in the energy-integrated response. Given that x-ray scattering cannot distinguish between static and dynamic contributions, neutron scattering measurements with energy discrimination and *in situ* strain will be needed to shed further light on this.

In contrast to the LTO features, the rod-like scattering at the nominally forbidden $(0\ \bar{3}\ 4)$ reflection (Fig. 6) is indicative of planar correlations. The dynamic component is likely due to a relatively soft phonon mode known from previous work [48], but the quasistatic part might arise from stacking faults or dislocations. Screw dislocations are an intriguing possibility, since they could generically appear due to the fact that the seed crystal and polycrystalline feed rod counter-rotate during crystal growth with the floating-zone technique. A network of screw dislocations might have interesting consequences for both structural and electronic properties, given the chiral nature of such dislocations. Targeted measurements of crystals grown under different conditions could shed further light on this unexplored issue. Yet from the present scattering measurements it cannot be conclusively determined if the observed fluctuations originate from defects or if they are a bulk effect. In the latter case, they could be due to a distortion that preserves the unit cell volume. One possible scenario is a trapezoidal distortion of oxygen atoms within the Cu-O planes. Such a distortion preserves the unit cell volume and would result in scattering at the $(0\ \bar{3}\ 4)$ and equivalent positions (see Appendix A).

The unusual LTO fluctuations were previously interpreted [15] as evidence for an underlying structural rare-region effect within Landau-Ginzburg-Wilson theory [49]. In particular, both the exponential scaling and power-law dependence of the characteristic length on absolute temperature are in qualitative agreement with rare-region theory. Similar to the findings in ref. [15], the exponent of the effective width in Fig. 4 is lower than the theoretically expected value of 1/2, but this might be due to the possible first-order nature of the LTO transition: the effective mean-field transition temperature would then be below $T_{LTO}$, which would lead to an apparent increase of the exponent in Fig. 4. Furthermore, close similarities were noted between the LTO fluctuations and the emergence of superconductivity on cooling [11-14], which appears to be dominated by geometric rather than thermal fluctuations, and with Lifshitz tails in the density of states of semiconductors in the well-known optical Urbach effect [50]. Since the LTO instability is non-universal and only appears in a few cuprate families, it was argued that there exists underlying, universal structural inhomogeneity that manifests itself *via* a coupling to structural and electronic order parameters [15]. The present work provides additional support for this hypothesis, most importantly from the uniaxial-stress measurement. The fact that the symmetry-breaking strain along [1 1 0] weakens the LTO diffuse intensity above $T_{LTO}$, but does not significantly change the temperature dependence, suggests that the latter is caused by an underlying phenomenon that is unrelated to the LTO instability and rather insensitive to stress, at least in the studied range. Moreover, our finding that the unusual exponential behavior and scaling persist in undoped $La_2CuO_4$ up to the highest measured temperatures suggests that this hidden inhomogeneity is formed during the crystal growth and independent of point disorder such as Sr/La substitution. Interestingly, the relevant energy scale of ~ 100 meV is therefore comparable to important electronic scales, *e.g.* the superconducting gap and several pseudogap scales [22,23], and it is possible that an interplay between structural and electronic degrees of freedom determines the energy scale of the inhomogeneity. While the nature of the inhomogeneity cannot be determined within the present study, we note that short-range structural correlations involving atomic displacements perpendicular to the $CuO_2$ planes have recently been observed in the simple-tetragonal cuprate $HgBa_2CuO_{4+\delta}$, with displacements of atoms in the ionic layer between the planes playing an important role [10]. It is a distinct possibility that such correlations are also present in LSCO and give rise to the observed LTO fluctuation behavior. Since octahedral rotations involve apical oxygen, and to a lesser extent lanthanum atoms, the out-of-plane correlations would couple to the LTO distortions. They would also be nearly insensitive to in-plane uniaxial strain, at least in the limited range studied here. Yet we have not observed unequivocal signatures of such correlations in LSCO diffuse x-ray and neutron scattering data due to the strong signals from octahedral rotations, and the latter also impede the use of 3D-ΔPDF analysis that has been successfully employed in $HgBa_2CuO_{4+\delta}$ and other correlated-electron systems [51]. More detailed measurements are thus needed to establish the presence and strength of such distortions in La-based cuprates.

The existence of intrinsic inhomogeneity could have far-reaching consequences in the study of complex oxides and other materials with strong electronic correlations. In general, if the local structure is significantly different than the average structure on length scales that are relevant for electronic physics, strong nonperturbative effects should be expected [52]. Both the normal-state mean-free paths and superconducting coherence lengths are short in the cuprates and other high-$T_c$ superconductors, such as the bismuthates [53] and iron-based systems [54], which implies that nanoscale inhomogeneity should play an important role. In addition, the possibility of scale-free inhomogeneity has been suggested: scattering experiments uncovered evidence for a fractal distribution of interstitial oxygen atoms in oxygenated La-based cuprates [9] and, as noted, different families of oxide superconductors show similar superconducting fluctuation regimes that are likely underpinned by inhomogeneity [14], despite large differences in superconducting coherence lengths [55-57]. The structural fluctuations studied here show close similarities to the universal superconducting fluctuations [11-14], which suggests a common origin. It is also an intriguing possibility that differing crystal growth conditions result in subtle variations in the local structural landscape that might have a nontrivial effect on their electronic and structural properties. As an example, the structural, magnetic, and electronic properties of the iridates $Sr_2IrO_4$ and $Sr_3Ir_4O_{10}$ can be dramatically altered *via* the application of a magnetic field during crystal growth [58]. If an analogous sensitivity to growth conditions is present in the cuprates, this would open a large parameter space to engineer their properties and have direct consequences for the comparison of measurements on crystals prepared with different methods. For example, it would complicate the comparison between thin-films and bulk materials, which have been observed to exhibit drastically different superfluid densities as superconductivity disappears on the overdoped side of the phase diagram [59]. Such variation with growth method in clean (low-impurity) crystals would suggest that these materials are not in thermodynamic equilibrium. Instead, this would imply that a local minimum in the free energy is realized by a particular microscopic structural configuration. Tuning the parameters during or after crystal growth might drive these systems to new local minima, or perhaps a global minimum, which would further change/enhance their electronic and structural properties.

In summary, we have investigated symmetry-lowering nanoscale structural fluctuations in the archetypal cuprate La$_{2-x}$Sr$_x$CuO$_4$. First, we used neutron and x-ray scattering to extend previous measurements of LTO fluctuations in the HTT phase to additional doping levels and temperatures, and we utilized a novel pneumatic strain cell to investigate the effect of in-plane stress along [1 1 0]. These measurements revealed that LTO fluctuations and their unusual scaling behavior persist to the highest measured temperatures, approaching a significant fraction of the melt temperature, which indicates that the underlying inhomogeneity forms during crystal growth. Moreover, the insensitivity of the response to uniaxial strain, as revealed from the temperature dependence, suggests that this underlying inhomogeneity is insensitive to in-plane stress. We also used diffuse neutron scattering to search for fluctuations at the forbidden $(\bar{3}\ 0\ 4)_{\text{HTT}}$ reflection. The observed elastic diffuse scattering indicates that the local structural symmetry is lower than the average orthorhombic symmetry.

**Appendix A**

*Structure-factor calculation*

In order to understand which type of distortion might lead to scattering intensity at the forbidden integer positions, we performed simple structure-factor calculations using the form

$$S(h,k,l) = \left| \sum_{j=1}^{N} f_j e^{-i(hx_j + ky_j + lz_j)} \right|^2$$

where $f_j$ and $(x_j, y_j, z_j)$ are the atomic form factor and coordinates of the $j^{th}$ atom in the unit cell. The atomic coordinates were taken as [60]

$$La_1 = \begin{pmatrix} 0.5 \\ 0.5 \\ 0.1391 \end{pmatrix}, \quad La_2 = \begin{pmatrix} 0 \\ 0 \\ 0.3609 \end{pmatrix}, \quad La_3 = \begin{pmatrix} 0 \\ 0 \\ 0.6391 \end{pmatrix}, \quad La_4 = \begin{pmatrix} 0.5 \\ 0.5 \\ 0.8609 \end{pmatrix}$$

$$Cu_1 = \begin{pmatrix} 0 \\ 0 \\ 0 \end{pmatrix}, \quad Cu_2 = \begin{pmatrix} 0.5 \\ 0.5 \\ 0.5 \end{pmatrix}$$

$$O_1 = \begin{pmatrix} 0.5 - r \\ -r \\ 0 \end{pmatrix}, \quad O_2 = \begin{pmatrix} r \\ 0.5 + r \\ 0 \end{pmatrix}, \quad O_3 = \begin{pmatrix} 0 \\ 0 \\ 0.184 \end{pmatrix}, \quad O_4 = \begin{pmatrix} 0.5 \\ 0.5 \\ 0.316 \end{pmatrix}$$

$$O_5 = \begin{pmatrix} 0.5 \\ 0.5 \\ 0.684 \end{pmatrix}, \quad O_6 = \begin{pmatrix} 0.5 - r \\ -r \\ 0.5 \end{pmatrix}, \quad O_7 = \begin{pmatrix} r \\ 0.5 + r \\ 0.5 \end{pmatrix}, \quad O_8 = \begin{pmatrix} 0 \\ 0 \\ 0.816 \end{pmatrix}$$

where $r = 0$ corresponds to the standard tetragonal unit cell, and $r > 0$ introduces a trapezoidal distortion of the oxygen atoms in the CuO$_2$ planes. The structure factor becomes nonzero at nominally forbidden integer positions in reciprocal space, including $(0\,\bar{3}\,4)$, for $r > 0$.

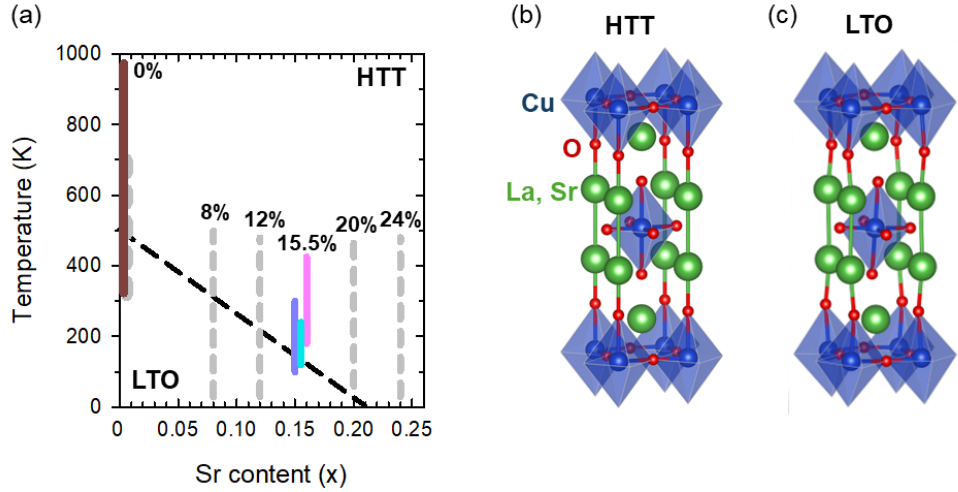

Fig. 1. (a) Schematic structural phase diagram of LSCO. The black dashed line indicates the transition at $T_{\text{LTO}}(x)$ that separates the high-temperature tetragonal (HTT) and low-temperature orthorhombic (LTO) phases [16]. Hole doping levels and temperatures at which the LTO correlations were studied in the present work are indicated by solid colored lines: brown ($x = 0$) and pink ($x = 0.155$) correspond to neutron scattering data; light blue ($x = 0.155$, measured without strain) and light purple ($x = 0.155$, measured with strain) correspond to x-ray scattering data. Dashed gray lines correspond to data from ref. [15]. The undoped system is a Mott insulator; superconductivity occurs in the approximate range $0.05 < x < 0.30$, with an optimal $T_c$ of nearly 40 K at $x \approx 0.15$ [28]. (b) Schematic *average* structure in the HTT phase, in which long-range $CuO_6$ rotations are absent. (c) The LTO distortion corresponds to a staggered rotation of the $CuO_6$ octahedra about the in-plane diagonals $[1\,1\,0]$ or $[1\,\bar{1}\,0]$, associated with two types of macroscopic structural domains; short-range LTO correlations are found to persist in the HTT phase.

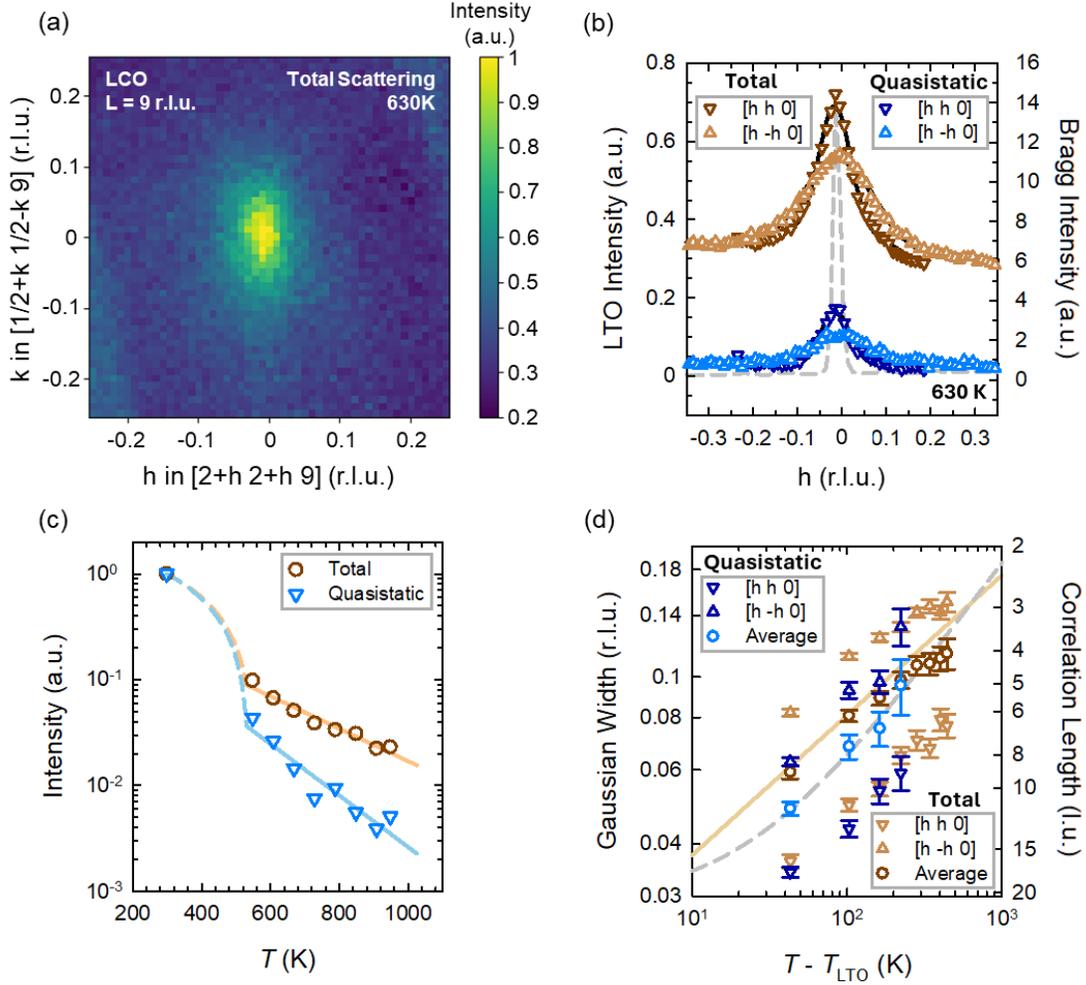

Fig. 2. Diffuse neutron scattering in the HTT phase of undoped $La_2CuO_4$ ($x = 0$), indicative of anisotropic short-range orthorhombic structural fluctuations. (a) Total scattering data showing the (5/2 3/2 9) peak in the $HK$9 plane at 630 K. The response is asymmetric along [1 1 0] and [1 $\bar{1}$ 0], consistent with previous observations for undoped/underdoped LSCO [15]. (b) Corresponding one-dimensional momentum cuts along [1 1 0] and [1 $\bar{1}$ 0]. As in ref. [15], the data were fit to a Gaussian profile with a third-order polynomial background (solid lines). The dashed gray line indicates the estimated momentum resolution, obtained by scanning along [1 1 0] through the nearby (3 2 9) Bragg peak. (c) Temperature dependence of the (5/2 3/2 9) peak intensity. The total and quasistatic scattering exhibit exponential behavior up to the highest measured temperature of 970 K. The quasistatic intensity decreases faster; the solid lines are the result of fits, as discussed in the text. Dashed lines are guides to the eye and indicate the increased scattering in the long-range-ordered phase (see also Fig. 4(a)). (d) Gaussian widths and correlation lengths along [110] and [1$\bar{1}$0] obtained from the fits, shown on a log-log scale. Also shown are the averages of the fit results. For the quasistatic response, the relatively low signal-to-noise ratio prevented reliable fits above 750 K. The quasistatic length is somewhat larger than that obtained from the total scattering and appears to increase at a higher rate upon approaching $T_{LTO}$. The solid line is the result of a power-law fit with exponent 1/3, whereas the grey dashed line assumes an effective $T'_{LTO} = T_{LTO} - 25$ K with the expected mean-field exponent of 1/2, which indicates that the data are consistent with a weakly first-order transition [15].

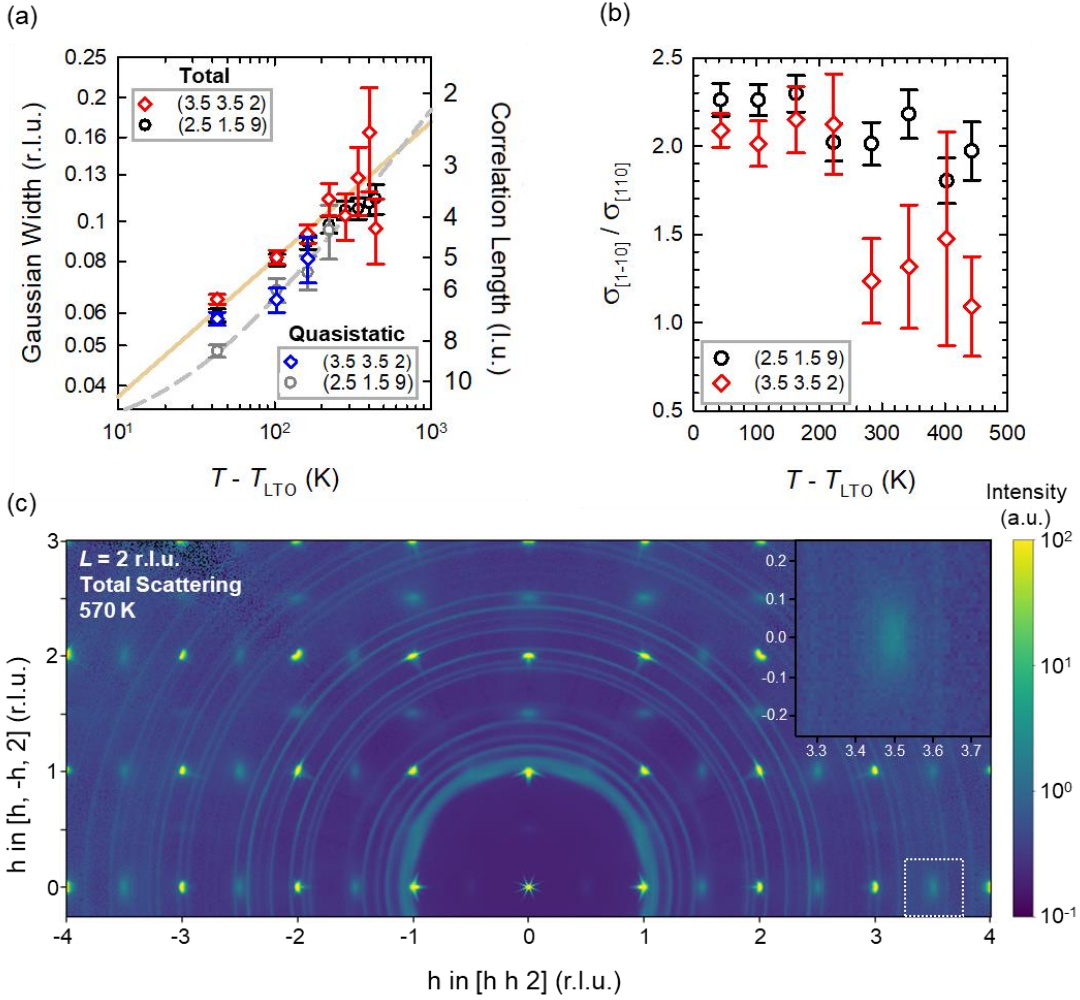

Fig. 3. Asymmetric LTO fluctuations in undoped $La_2CuO_4$ ($x = 0$) obtained from neutron scattering. The result obtained from measurements of the (5/2 3/2 9) peak (see Fig. 2) are compared with those for the (7/2 7/2 2) peak. The momentum transfer in the latter case is nearly parallel to [1 1 0], which corresponds to one of the two LTO tilt directions (Fig. 1(b)). (a) Gaussian width and correlation length for the total and quasistatic scattering channels, shown on a log-log scale. Results were extracted from cuts along [1 1 0] and [1 $\bar{1}$ 0] and then averaged; quasistatic scattering results are only shown up to 750 K for the (5/2 3/2 9) peak and 690 K for the (7/2 7/2 2) peak, since the relatively low signal at higher temperatures prevented a reliable analysis. The data for the two Brillouin zones are overall consistent with each other. (b) Temperature dependence of the ratio of the Gaussian widths, $\sigma_{[1\bar{1}0]}/\sigma_{[110]}$. (c) Total scattering data in the $L = 2$ plane at 570 K. The LTO peaks are elongated along either [1 1 0] or [1 $\bar{1}$ 0]. Powder rings from Al and Pt are clearly visible. The (7/2 7/2 2) peak is indicated by the dashed square; the inset shows a magnified view.

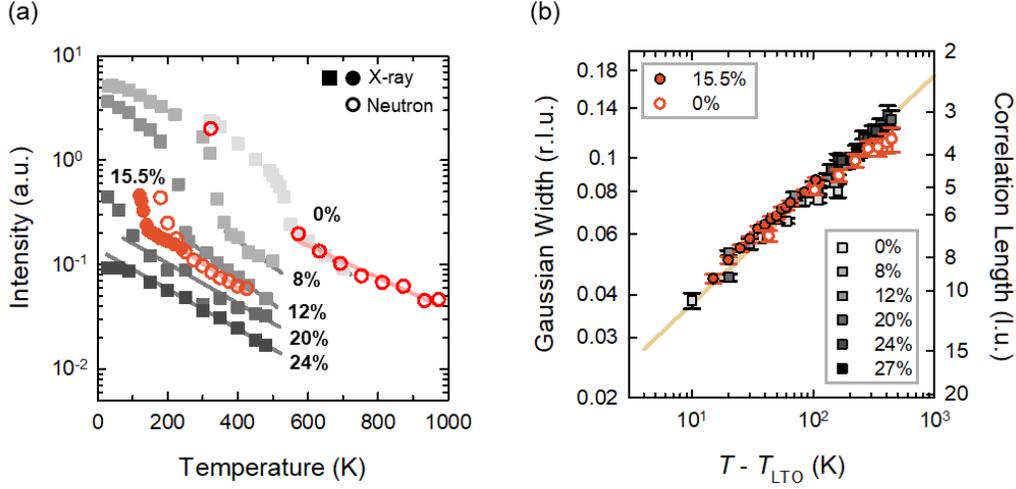

Fig. 4. Temperature dependence of (a) LTO scattering intensity and (b) correlation length for current and prior [15] x-ray (solid symbols) and neutron (open symbols) scattering results, shown on semi-log and log-log scales, respectively. The squares indicate the prior x-ray results [15]. The data obtained in the present work are indicated by red circles; the intensities in (a) were scaled to match the prior results. For $x = 0$, the total scattering neutron result is shown. The characteristic temperature scale of the exponential intensity decrease is essentially the same for the neutron and x-ray results and, as previously observed [15], nearly independent of doping. The neutron scattering results for the undoped ($x = 0$) and optimally doped ($x = 0.155$) samples were obtained at ORNL with the CORELLI time-of-flight and the HB-3 triple-axis instrument, respectively. The relatively broad structural transition in the latter case is the result of macroscopic Sr/La inhomogeneity in the large ~6 g crystal that was studied. Correlation lengths for this sample are not shown in (b) because the broad transition resulted in a poorly defined bulk $T_{LTO}$. Note that the results in (b) are plotted *vs.* relative temperature, $T - T_{LTO}$, and that the solid line indicates the same power-law behavior as in Figs. 2(d) and 3(a).

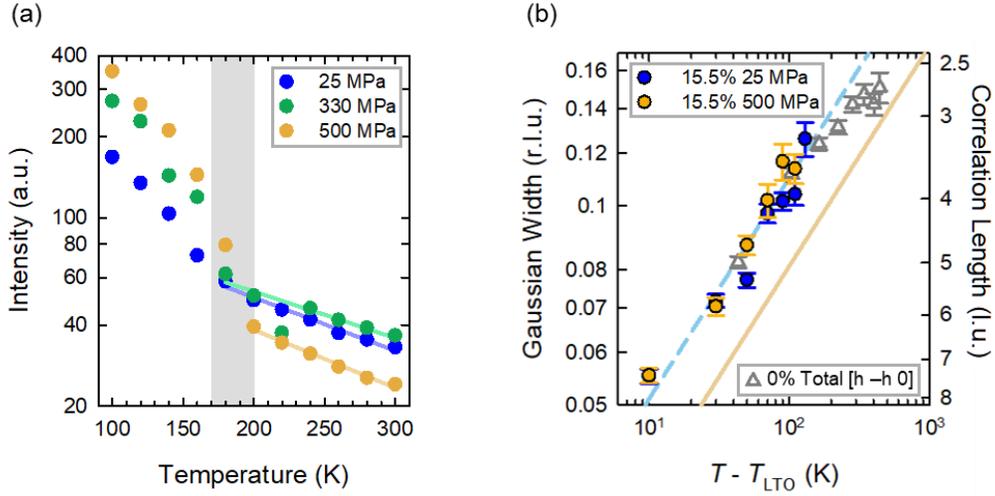

Fig. 5. (a) Temperature dependence of the diffuse x-ray scattering intensity for a $x = 0.155$ sample with *in situ* uniaxial strain along [1 1 0]. Stresses of 25, 330, and 500 MPa were applied above the structural transition temperature, $T_{LTO}$. The highest stress value corresponds to a strain of about 0.2%. Multiple reflections were measured and averaged (see Experimental Methods). At the highest stress value, the intensity of the LTO fluctuations in the HTT phase is seen to decrease considerably, whereas the Bragg scattering in the ordered phase is enhanced. The characteristic exponential temperature scale ($T_0$) remains essentially unchanged, whereas at 500 MPa, there appears to be a slight (~ 20 K) increase of $T_{LTO}$. (b) Correlation length and Gaussian width temperature dependences for the ~25 MPa and 500 MPa strain-cell data. The results are indistinguishable. The dashed blue line is a power-law fit with exponent 1/3, consistent with the unstrained results. However, the extracted correlation lengths for the strain-cell data are systematically smaller than those of unstrained samples (the solid yellow line corresponds to those in Figs. 2d, 3a, 4b), and consistent with the total scattering correlation length for undoped LCO ($x = 0$) along [1$\bar{1}$0] (gray triangles, see also Fig. 2d).

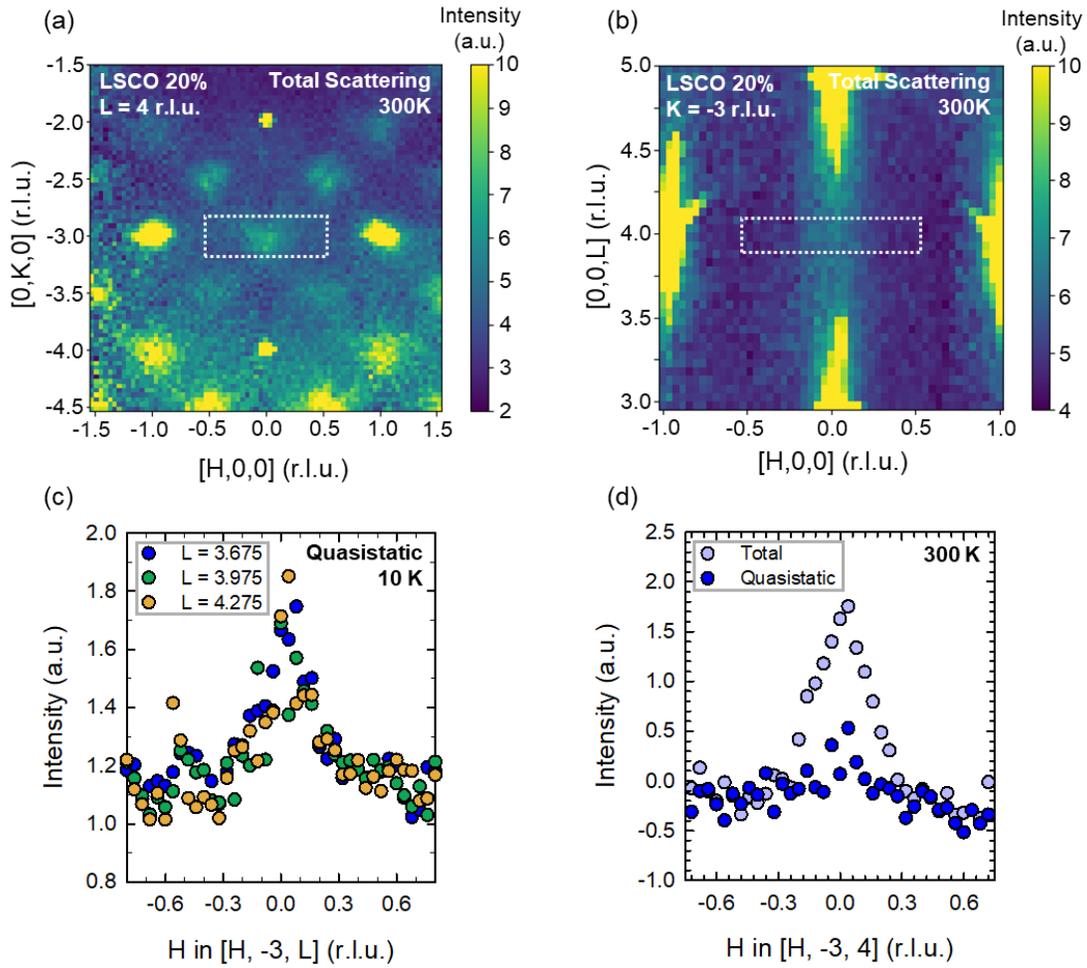

Fig. 6. (a) Total diffuse neutron scattering response in the (*H K* 4) plane for a *x* = 0.20 sample, obtained with the CORELLI instrument at 300 K. Incommensurate peaks corresponding to short-range LTO correlations are seen at the half-integer positions, as reported previously [15]. In addition, a clear diffuse signal is observed at nominally forbidden reflections such as (0 $\bar{3}$ 4). (b) The corresponding data in the (*H* $\bar{3}$ *L*) plane indicate rods of scattering along *L*, and hence two-dimensional correlations. (c) Line cuts through the rod-like feature in the quasistatic channel at 10 K. Little variation is observed along *L*. (d) Comparison of the total and quasistatic scattering. The response is primarily inelastic, although there is an elastic component indicating that static distortions are present. The dashed boxes in (a,b) indicate the approximately ± 0.1 r.l.u. integration ranges for the data shown in (d).